\documentclass[12pt]{article}
\usepackage{amssymb}
\usepackage{epsfig}

\newcommand{\bq}{\begin{equation}}
\newcommand{\eq}{\end{equation}}
\newcommand{\bqa}{\begin{eqnarray}}
\newcommand{\eqa}{\end{eqnarray}}
\newcommand{\ben}{\begin{enumerate}}
\newcommand{\een}{\end{enumerate}}
\newcommand{\bc}{\begin{center}}
\newcommand{\ec}{\end{center}}
\newcommand{\sss}{\tilde{s}_l}
\newcommand{\ccc}{\tilde{c}_l}

\def\pr#1#2#3{ Phys. Rev. ${\bf{#1}}$ (#2) #3}
\def\prl#1#2#3{ Phys. Rev. Lett. ${\bf{#1}}$ (#2) #3}
\def\pl#1#2#3{ Phys. Lett. ${\bf{#1}}$ (#2) #3}
\def\prep#1#2#3{ Phys. Rep. ${\bf{#1}}$ (#2) #3}

\def\ijmp#1#2#3{ Int. J. Mod. Phys. ${\bf{#1}}$ (#2) #3}

\def\epjc#1#2#3{ Eur. Phys. Jour. ${\bf{#1}}$ (#2) #3}
 
\def\etal{{\it et.al.\/}}

\global\nulldelimiterspace = 0pt

\def\L{ {\cal L }}
\def\O{ {\cal O }}

\begin{document}
\thispagestyle{empty}
\begin {flushleft}
PM/99-38\\
November 1999\\
\end{flushleft}

\vspace*{2cm}

%---------------------titre ---------------------------------------
\hspace*{-0.5cm}
\begin{center}
{\Large {\bf  Bounds on universal new physics effects from
fermion-antifermion production at LEP2}}\\ 
\hspace*{-0.5cm}
\vspace{1.cm} \\{\large M. Beccaria$^{a,b}$ F.M.
Renard$^c$,
S. Spagnolo$^d$ and C. Verzegnassi$^{e,b}$}\hspace{2.2cm}\null
\vspace {0.5cm} \\
\begin{center}
$^a$Dipartimento di Fisica, Universit\`a di 
Lecce \\
Via Arnesano, 73100 Lecce, Italy.\\
\vspace{0.2cm}  
$^b$INFN, Sezione di Lecce\\
Via Arnesano, 73100 Lecce, Italy.\\
\vspace{0.2cm} 
$^c$ Physique
Math\'{e}matique et Th\'{e}orique, UMR 5825\\
Universit\'{e} Montpellier
II,  F-34095 Montpellier Cedex 5.\hspace{2.2cm}\\
\vspace{0.2cm} 
$^d$ 
Rutherford Appleton Laboratory - Particle Physics Department \\
Chilton, Didcot, Oxfordshire OX11 0QX\\
\vspace{0.2cm}
$^e$
Dipartimento di Fisica Teorica, Universit\`a di Trieste, \\
Strada Costiera
 14, Miramare (Trieste)\\
\end{center}

\vspace*{3cm}
{\bf Abstract}\hspace{2.2cm}\null
\end{center}
\hspace*{-1.2cm}
\begin{minipage}[b]{16cm}
We consider lepton-antilepton annihilation into a fermion-antifermion
pair at variable c.m. energy. We propose for this process a simple
parametrization of the virtual effects of the most general model of new
physics of \underline{universal} type. This parametrization is based on
a recent approach, that uses the experimental results of LEP1, SLC as
theoretical input. It introduces \underline{three} functions whose
energy dependence is argued to be smooth and, in first approximation,
negligible. A couple of representative models of new physics 
are considered, as a
support of the previous claim. Explicit bounds are then derived for
this type of new physics from the available LEP2 data, and a discussion
is given of the relevance in this respect of the different experimental
measurements. The method is then extended to treat the case of two
particularly simple models of {\it non universal} type, for which it
is possible to draw analogous conclusions.
\end{minipage} 

\setcounter{footnote}{0} 
\clearpage
\newpage 
  
\hoffset=-1.46truecm
\voffset=-2.8truecm
\textwidth 16cm
\textheight 22cm
\setlength{\topmargin}{1.5cm}

\section{Introduction.} 

One of the most interesting consequences of the high precision results
obtained at LEP1, SLC \cite{LEPcombi} has been the determination
of bounds on the virtual effects of models of new physics. This had led
in some cases to drastic conclusions from which very useful indications
about the possible "surviving" models have been drawn.\par
One fundamental step in this process of derivation of bounds has been
the introduction of "simple" parametrizations of the new physics effects.
In this respect, we feel, both the original idea of Peskin and Takeuchi
of introducing two ($S$, $T$) parameters \cite{Peskin}, and the later
more rigorous proposal of Altarelli and Barbieri, leading to the
definition of $\epsilon_1$, $\epsilon_3$ \cite{AB}, 
have played a fundamental
role for any meaningful investigation of new physics effects at the $Z$
resonance \cite{ZNP}.\par
Strictly speaking, the previous two-parameter (e.g. $\epsilon_1$,
$\epsilon_3$) description of new physics is only possible if the latter
is of \underline{universal} type, i.e. independent of the flavour of
the final fermion-antifermion pair. At the $Z$ peak, the remarkable
simplification occurs that box diagrams can be ignored, being
kinematically suppressed. Therefore, the only possible non universal
effects may come from the final vertices (the most remarkable example
being that of the $b\bar b$ vertex). When these are carefully taken
into account, the theoretical description of new physics becomes less
simple, but still "relatively" straightforward. This is also due to
the fact that, at the $Z$ peak, the s-channel photon exchange, with
all its new
physics content, can be safely neglected.\par
An important question that arises in a natural way is that of whether
the previous, reasonably simple, picture will continue to be valid at
the higher energy electron-positron accelerators, in particular at
LEP2, but also, looking at a more distant future, at a linear
electron-positron collider (LC)\cite{LC} or at a muon-muon
collider\cite{mumu}. Here all the previous simplifications occuring on
the $Z$ resonance are no longer valid. In particular, both the photon
exchange with its universal and non universal components and the non
universal box diagrams at one loop cannot be now neglected \cite{box}.
Also, another technical advantage of the experiments at the $Z$ peak
is lost: present and future measurements in the two fermion production
process will be spread out over a range of different center of mass
energies, leading in general to a diluted experimental precision in
the investigation of small, $q^2$ dependent, 
deviations from the Standard 
Model predictions.
This certainly gives the impression that the
theoretical description of new physics effects might be, in general,
much more cumbersome than at the $Z$ peak.\par
The aim of this short paper is that of showing that, for what concerns
the process of final fermion-antifermion production, the situation is
still promising if one restricts the investigation to effects of new
physics of \underline{universal} type. Here, under "reasonable"
assumptions, a situation that appears as the immediate
generalization of that met at the $Z$ peak, with only \underline{three}
parameters rather than two (the extra one being an obvious relic of the
photon exchange contribution), will be proposed and justified with
some examples.\par
The starting point of our analysis will be the use of a theoretical
description of the process of electron-positron annihilation into a
charged fermion-antifermion pair (with the exclusion for the moment of
Bhabha scattering that will be treated in a separate forthcoming paper)
that was called the "$Z$-peak subtracted" approach. This has been
exhaustively illustrated and discussed in previous references
\cite{Zsub} and here we shall neither repeat its motivations nor recall
the relevant features. The only technical detail that we shall use is
the expression of the differential \underline{unpolarized} cross
section at variable squared c.m. energy $q^2$ and scattering angle
$\theta$; for our purposes, it will be convenient to write it in the
following form ($f$ denotes the final fermion of the considered
process, that will be either a lepton ($\mu$, $\tau$) or a quark, more
precisely a "light" ($u$, $d$, $s$, $c$, $b$) one); $\ell$ indicates the
{\it initial} lepton, which is for the purposes of this paper an electron):

\bq
{d\sigma_{lf}\over dcos\theta}={4\pi\over3}N_fq^2\{{3\over8}
(1+cos^2\theta)U_{11}+{3\over4}cos\theta
U_{12}\}
\label{sig}\eq
\noindent
where
\bqa  
U_{11}=&&
{\alpha^2(0)Q^2_f\over
q^4}[1+2\tilde{\Delta}_{\alpha, lf}(q^2,\theta)]
\nonumber\\
&&+2[{\alpha(0)|Q_f|}]{q^2-M^2_Z\over
q^2((q^2-M^2_Z)^2+M^2_Z\Gamma^2_Z)}[{3\Gamma_l\over
M_Z}]^{1/2}[{3\Gamma_f\over N_f M_Z}]^{1/2}
{\tilde{v}_l \tilde{v}_f\over
(1+\tilde{v}^2_l)^{1/2}(1+\tilde{v}^2_f)^{1/2}}\nonumber\\
&&\times[1+
\tilde{\Delta}_{\alpha,lf}(q^2,\theta) -R_{lf}(q^2,\theta)
-4\tilde s_l\tilde c_l
\{{1\over \tilde{v}_l}V_{lf}^{\gamma Z}(q^2,\theta)
+{|Q_f|\over\tilde{v}_f}
V_{lf}^{Z\gamma}(q^2,\theta)\}]\nonumber\\ 
&&+{[{3\Gamma_l\over
M_Z}][{3\Gamma_f\over N_f M_Z}]\over(q^2-M^2_Z)^2+M^2_Z\Gamma^2_Z}
\nonumber\\
&&\times[1-2R_{lf}(q^2,\theta)
-8\tilde s_l\tilde c_l\{{\tilde{v}_l\over1+\tilde{v}^2_l}V_{lf}^{\gamma
Z}(q^2,\theta)+{\tilde{v}_f|Q_f|\over(1+\tilde{v}^2_f)}
V_{lf}^{Z\gamma}(q^2,\theta)\}]
\label{U11pro}
\eqa
\bqa
U_{12}=&& 2[{\alpha(0)|Q_f|}]{q^2-M^2_Z\over
q^2((q^2-M^2_Z)^2+M^2_Z\Gamma^2_Z)}
[{3\Gamma_l\over M_Z}]^{1/2}[{3\Gamma_f\over N_f
M_Z}]^{1/2}{1\over(1+\tilde{v}^2_l)^{1/2}(1+\tilde{v}^2_f)^{1/2}}
\nonumber\\
&&\times[1+
\tilde{\Delta}_{\alpha,lf}(q^2,\theta)
-R_{lf}(q^2,\theta)]\nonumber\\
&&+{[{3\Gamma_l\over M_Z}][{3\Gamma_f\over N_f
M_Z}]\over(q^2-M^2_Z)^2+M^2_Z\Gamma^2_Z}
[{4\tilde{v}_l \tilde{v}_f\over(1+\tilde{v}^2_l)(1+\tilde{v}^2_f)}]
\nonumber\\
&&\times[1-2R_{lf}(q^2,\theta)-4\tilde s_l\tilde c_l
\{{1\over \tilde{v}_l}V_{lf}^{\gamma Z}(q^2,\theta)+{|Q_f|\over\tilde{v}_f}
V_{lf}^{Z\gamma}(q^2,\theta)\}]
\label{U12pro}
\eqa

In the previous equations, $\Gamma_{l,f}$ are the partial $Z$ widths
into a {\it final} lepton ($l\bar l$) or into a general fermion($f\bar f$)
 pair; $\tilde s^2_l=1-\tilde c^2_l$ is 
the
effective weak angle $sin^2\theta_{W,eff}$ (for simplicity we neglect
here very small differences between parameters entering at the
one loop level); 
$\tilde v_l=1-4\tilde s^2_l$;
$\tilde v_f=1-4|Q_f|\tilde s^2_l$; 
$N_f$ is the colour factor including QCD
corrections.\par
In the SM, the four functions $\tilde\Delta_{\alpha,lf}$, 
$R_{lf}$, $V^{\gamma Z}_{lf}$,
$V^{Z\gamma}_{lf}$ that appear in the various expressions are
\underline{gauge-invariant} combinations of one-loop self-energies,
vertices and boxes. They determine the $\theta$-integrated expression
of the various unpolarized cross sections and asymmetries in a way that
it is immediate to derive for self-energies and vertices, and requires
a more complicated numerical calculation in the case of boxes, that are
generally $\theta$ dependent.\par
When one considers models of new physics (NP) of electroweak type, whose
virtual effects on the considered fermion-antifermion production
process do not add extra Lorentz structures to the scattering
amplitude at one loop, the previous description of Eqs.(\ref{sig})-
(\ref{U12pro}) still
remains valid, with the formal replacement

\bq
\tilde\Delta_{\alpha,lf}(q^2,\theta) \to
\tilde\Delta_{\alpha,lf}(q^2,\theta) 
+ \tilde\Delta^{NP}_{\alpha,lf}(q^2,\theta)
\label{DNP}
\eq
\noindent
(similarly for $R_{lf}$, $V^{\gamma Z}_{lf}$ and
$V^{Z\gamma}_{lf}$). This leads to a straightforward modification of the 
expressions of the various
observables where the effect of the extra model can be estimated, at
each $q^2$, once the expressions of the various functions
$\left(\tilde\Delta_{\alpha_{lf}}, R_{lf}, V^{\gamma Z}_{lf},
V^{Z\gamma}_{lf}\right)^{NP}$ have been given. 
One might then proceed
to the determination of bounds for the new physics effects, at each
$q^2$ value that was considered, after a procedure that implies the
angular integration.The latter will leave, in general, several 
{\it different} functions of $q^2$ generated by the different powers
of $\cos \theta$ that appear in the integrand.
A typical example of such a situation would be the
determination of the effects due to SUSY boxes, that were already
discussed in a special case, at the LEP2 energies, in a previous paper
\cite{SUSYbox}. For these cases, the determination of the bounds would
require a dedicated computational program, in agreement with the
expectation that was anticipated in this paper.\par
 A first simplification occurs for models of new physics that do not
have any $\theta$-dependence. In such a case, the extra contribution to
the observables can be very easily expressed in terms of four different
functions of $q^2$ only, to be indicated in this paper as
$\tilde\Delta^{NP}_{\alpha}(q^2)$, $R^{NP}(q^2)$, $V^{\gamma
Z,NP}(q^2)$,
$V^{Z\gamma,NP}(q^2)$. For what concerns these functions, it would become
possible to derive bounds for their \underline{values} at each
different $q^2$ that is experimentally relevant. This would require a
separate analysis for each different model of new physics type, that
would also depend on the details of the final state on which the four
functions might depend.\par
A second remarkable simplification occurs for models of new physics
whose effects are of \underline{universal} type (UNP). In such cases,
only \underline{three} functions of $q^2$ would remain in the
theoretical
expression, since  $V^{\gamma Z,NP}=V^{Z\gamma,NP}$ 
in this case \cite{Zsub}.
These functions will be called from now on 
$\tilde\Delta^{UNP}_{\alpha}(q^2)$, $R^{UNP}(q^2)$, $V^{UNP}(q^2)$.The
previous scenario is particularly appealing, not only because of the gain
in conceptual simplicity achieved through the reduction of the overall
number of unknown functions, but also because the measurements of fermion pair
production for different final flavours can be combined to determine the best 
fit values for the NP contributions, leading to bounds that will still be
$q^2$ dependent.\par 
A final and drastic simplification occurs if we introduce the physical
assumption that the models of new physics that are considered contain
an intrinsic scale that is essentially larger than the $q^2$ values at
which the measurements are performed. {\it In the $Z$-peak subtracted
approach}, this assumption can be exploited in a very useful way. In
fact, one of the characteristic features of this approach is that the
three functions $\Delta^{UNP}_{\alpha}(q^2)$, $R^{UNP}(q^2)$,
$V^{UNP}(q^2)$ must always \underline{vanish}, respectively, at the
$q^2$ values that correspond to the photon or to the $Z$ mass. In other
words, one has by construction:

\bq
\tilde\Delta^{UNP}_{\alpha}(q^2=0)=R^{UNP}(q^2=M^2_Z)=V^{UNP}(q^2=M^2_Z)
=0
\label{substrac}
\eq

As an immediated consequence of Eq.(\ref{substrac}), 
one is entitled to write:

\bq
R^{UNP}(q^2) = {(q^2-M^2_Z)\over M^2_Z}\ \delta_z
\label{RUNP}
\eq
\bq
V^{UNP}(q^2) = {(q^2-M^2_Z)\over M^2_Z}\ \delta_s
\label{VUNP}
\eq
\bq
\tilde\Delta_{\alpha}^{UNP}(q^2) = {q^2\over M^2_Z}\ \delta_{\gamma}
\label{DUNP}
\eq
\noindent
where $\delta_{z, s, \gamma}$ are dimensionless functions of
$q^2$. 
As indicated by the indices, $\delta_z$ refers to
the modification of the $Z$ coupling, $\delta_{\gamma}$ to the
modification of the photon coupling and $\delta_{s}$ to that of the
effective weak angle~\cite{Zsub,Log}.

For values of the intrinsic new physics scale sufficiently
larger than the c.m. energies at which the experimental searches are
performed, we can reasonably expect that the $q^2$ dependence of  
$\delta_{z,s,\gamma}$ is smooth. In this case, it will be reasonable to
approximate the $\delta_i$ by the lowest order terms in their
$q^2$ expansion. 
This procedure will
reduce the functions $\delta_{s,\gamma, z}$ to three constants, i.e.
the coefficients of the lowest $q^2$ power. This will correspond to
writing $\delta_i(q^2)\simeq\delta_i(0) \equiv \delta_i$ in all 
cases where $\delta_i(0)\neq 0$.

For the class of new physics models that meet all the previous
requirements (that, as we
shall explicitely show with some examples, is not empty) the effect 
on the various observables is rather simple.
We have only considered in this paper the case of
unpolarized initial beams and concentrated our attention on five
different observables, that are measured at LEP2. These are: the cross
section for muon (or tau) production $\sigma_{\mu}$; the related
forward-backward asymmetry $A_{FB,\mu}$; the cross section for five
"light" ($u$, $d$, $s$, $c$, $b$) quark production $\sigma_5$; the
cross section for ($b\bar b$) production $\sigma_b$ and the related
forward-backward asymmetry $A_{FB,b}$. In terms of the three parameters
$\delta_z$, $\delta_s$, $\delta_{\gamma}$ 
the effects of the most general
model of new physics of universal type can be written as follows (we
use systematically, for a given observable $\O_i$, the notation 
$\O_i=\O_i^{SM}[1+ d\O^{UNP}_i/\O_i^{SM}]$ while the  experimental
value and errors is denoted by $\O_i^{meas}\pm \sigma_i$):

\bqa
{d\sigma^{UNP}_{\mu}\over\sigma_{\mu}}= {2\over M^2_Z[
7.99(q^2-M^2_Z)^2+q^4]}&\{&\delta_{\gamma}~[7.99q^2(q^2-M^2_Z)^2]
-\delta_z~(q^2-M^2_Z)q^4\nonumber\\
&&-\delta_s~(q^2-M^2_Z)~[0.70q^2(q^2-M^2_Z)+0.25q^4]~\}
\label{dsigmu}\eqa
\bqa
{dA^{UNP}_{FB,\mu}\over A_{FB,\mu}}= {1\over M^2_Z[
7.99(q^2-M^2_Z)^2+q^4]}&\{&(\delta_{\gamma}q^2+\delta_z~(q^2-M^2_Z))~
[q^4-7.99(q^2-M^2_Z)^2\nonumber\\
&&+\delta_s~(q^2-M^2_Z)~[0.49q^4-0.18{q^6\over q^2-M^2_Z}]~\}
\label{dAFBmu}\eqa
\bqa
{d\sigma^{UNP}_{5}\over\sigma_{5}}&=& {1\over M^2_Z[
0.81q^4+0.06q^2(q^2-M^2_Z)+(q^2-M^2_Z)^2]}\times\nonumber\\
&&\{\delta_{\gamma}q^2~
[2(q^2-M^2_Z)^2+0.06q^2(q^2-M^2_Z)]
-\delta_z~(q^2-M^2_Z)~[0.06q^2(q^2-M^2_Z)+1.62q^4]\nonumber\\
&&-\delta_s~(q^2-M^2_Z)~[1.46q^2(q^2-M^2_Z)+0.89q^4]~\}
\label{dsig5}\eqa
\bqa
{d\sigma^{UNP}_{b}\over\sigma_{b}}&=& {1\over M^2_Z[
5.53(q^2-M^2_Z)^2+0.59q^2(q^2-M^2_Z)+9q^4]}\times\nonumber\\
&&\{\delta_{\gamma}q^2~
[5.53(q^2-M^2_Z)^2+0.30q^2(q^2-M^2_Z)]
-\delta_z~(q^2-M^2_Z)~[0.30q^2(q^2-M^2_Z)+9q^4]\nonumber\\
&&-\delta_s~(q^2-M^2_Z)~[6.99q^2(q^2-M^2_Z)+3.48q^4]~\}
\label{dsigb}\eqa
\bqa
{dA^{UNP}_{FB,b}\over A_{FB,b}}={d\sigma^{UNP}_{FB,b}\over\sigma_{FB,b}}
-{d\sigma^{UNP}_{b}\over\sigma_{b}}
\label{dAFBb}\eqa
\noindent
with\bqa
{d\sigma^{UNP}_{FB,b}\over\sigma_{FB,b}}&=& {1\over M^2_Z[
1.93q^2(q^2-M^2_Z)+0.21q^4]}\{\delta_{\gamma}q^2~
[1.93q^2(q^2-M^2_Z)]\nonumber\\
&&-\delta_z~(q^2-M^2_Z)~[1.93q^2(q^2-M^2_Z)+0.41q^4]
-\delta_s~(q^2-M^2_Z)~[4.88q^4]~\}
\label{dsigFBb}\eqa
\noindent
where, in the various kinematical coefficients, the values of the various
$Z$ peak inputs that enter Eqs.(\ref{dsigmu}-\ref{dsigFBb}) 
have been taken by the most recent
experimental review \cite{LEPcombi}.\par

In order to demonstrate the predictivity of the proposed parametrization
of virtual NP
effects, we used a set of preliminary results \cite{LEP2} from the LEP
collaborations obtained with the largest available sample of data collected
at a single center of mass energy, i.e. the data of the 1998 LEP run at
$\sqrt{s} = 189$~GeV.  
The most probable values of the $\delta$ parameters are obtained from the
minimization of a $\chi^2$ defined in terms of the deviations of the
present measurements with respect to the SM predictions and of the residual 
NP contributions to the theoretical expressions of the observables: 
$$
\chi^2 = \sum_{i=1}^{N} \left(
 \frac{{\cal O}^{meas}_i-{\cal O}^{\rm SM}_i-d{\cal O}^{\rm
     UNP}_i}{\sigma_i}
\right)^2
$$
where the index $i$ corresponds to a specific observable measured in the
two fermion process and the theoretical deviations with respect to 
the SM are expressed as a function of the $\delta$ parameters according to
Eqs. (\ref{dsigmu}-\ref{dsigFBb}). 
Since the experimental precision of the measurements determines 
the ultimate constraints on the UNP parameters, the present accuracy on
$\rm R_b$ and $\rm A_{FB,b}$ makes their contribution to the overall
$\chi^2$ negligible with respect to the leptonic cross
sections and asymmetries and to the $q\bar{q}$ production. 
Therefore, although they will carry useful information in a final analysis, 
where the whole LEP2 data set can be exploited, we will not use them 
in the following discussions.\par 
Fig.\ref{189_4exp} shows that the measurements of 
$\sigma_\mu, \:\sigma_\tau, \: \sigma_5, \: 
{\rm A_{FB\mu}}$ and ${\rm A_{FB\tau}}$ from  the four LEP experiments 
\cite{LEP2} lead to simultaneous determinations of 
the $\delta$ parameters which appear
statistically consistent among them and in agreement, within the
experimental uncertainty, with the SM
expectations. 
There is no need to stress here the relevance of a combination
of the results of the different experiments, which, accounting for the
possible correlations of some sources of systematic experimental 
uncertainties, allows to improve the sensitivity of 
this kind of analysis
to potential hints of NP.  
The results of a first combination \cite{LEP2comb}, which 
is far from being trivial due to 
the different signal definitions adopted by each experiment, 
have been used
in the following in order to obtain a 
determination of $\delta_z$, $\delta_s$ and $\delta_{\gamma}$ from the 
whole data sample collected at $\sqrt{s}=189$~GeV.  
Taking in mind the premature nature of this combination and the
preliminary results used as inputs, the
resulting determination, shown in 
Fig.\ref{189_4exp} is mainly intended to evidentiate the size of
the ultimate constraints on our general parameterization of UNP effects 
allowed by the use of the entire set of experimental results obtained 
at $\sqrt{s} = 189$~GeV, rather than to give the final bounds on 
the $\delta$ parameters. 

The hypothesis of smooth $q^2$ dependence of the NP effects 
can be exploited
to further improve these constraints with the measurements of the two
fermions observables at other center of mass energies. Table \ref{tab1}
shows the size of the 95\% C.L. exclusion regions
obtained by a simultaneous minimization of the $\chi^2$ with 
respect to the
three parameters when, for each observable, the experimental precision 
of the combination \cite{LEP2comb} of the data from the four 
LEP2 experiments at
$\sqrt{s}=189$~GeV and $\sqrt{s}=183$~GeV is assumed. 
Since the sensitivity of the two fermion observables to $\delta_i$
increases in our approach with the $q^2$ of the process, further significant 
improvement of
these constraints may be envisaged in the near future from the data
collected at higher center of mass energies in the last two years of 
LEP2 operation.\par
Before attempting an estimate of the final combined LEP2 sensitivity to 
$\delta_i$, we illustrate a couple of examples of New Physics 
of typical scales
$\Lambda$ larger than the electroweak one, whose residual 
virtual effects 
at LEP2 energies can be parameterized according to the scheme 
proposed in 
this paper. In such examples the specificity of the models will also
entail a further reduction of the number of free parameters, 
thus allowing
for more stringent constraints to be derived from the 
experimental data.\par
The first case that we considered was that of Anomalous Gauge Couplings
(AGC). We used 
the framework of Ref.~\cite{agc} in which the effective
Lagrangian is constructed with dimension six operators respecting
$SU(2)\times U(1)$ and
$CP$ invariance. As shown in Ref.~\cite{agczsub}, 
only two parameters ($f_{DW}$
and $f_{DB}$) survive in the Z-peak subtracted approach.
The explicit expression of the UNP contribution to
$\delta_z$, $\delta_s$ and $\delta_{\gamma}$ are 
\bq
\delta_z = 8\pi\alpha \frac{M_Z^2}{\Lambda^2} 
\left(\frac{\ccc^2}{\sss^2} f_{DW} + \frac{\sss^2}{\ccc^2} 
f_{DB}\right) ,
\label{AGC1}\eq
\bq
\delta_s = 8\pi\alpha \frac{M_Z^2}{\Lambda^2} 
\left(\frac \ccc \sss f_{DW} -
\frac \sss \ccc f_{DB}\right) ,
\label{AGC2}\eq
\bq
\delta_{\gamma} =
-8\pi\alpha\frac{M_Z^2}{\Lambda^2}\left(f_{DW}+f_{DB}\right) ,
\label{AGC3}\eq
They satisfy the linear constraint:
\bq
\delta_z -\frac{1-2\sss^2}{\sss\ccc} \delta_s+\delta_{\gamma} = 0 .
\label{AGCrel}\eq

The second considered model was one of Technicolour type 
with two families of
strongly coupled
resonances ($V$ and $A$) \cite{tc}. The typical UNP 
parameters are the two
ratios $F_A/M_A$ and $F_V/M_V$ where $F_{A, V}$ and 
$M_{A, V}$ are the
couplings and the masses (that in this case play the role of the new
physics scale $\Lambda^{TC} >> q^2$) of the {\it lightest} axial 
and vector resonances.
The contribution to $\delta_z$, $\delta_s$ and $\delta_{\gamma}$ are
\bq
\delta_z = M_Z^2 \frac{\pi\alpha }{\sss^2\ccc^2} \left(
(1-2\sss^2)^2 \frac{F_V^2}{M_V^4}+\frac{F_A^2}{M_A^4}\right) ,
\label{TC1}\eq
\bq
\delta_s = M_Z^2 \frac{2\pi\alpha}{\sss\ccc} (1-2\sss^2) 
\frac{F_V^2}{M_V^4} ,
\label{TC2}\eq
\bq
\delta_{\gamma} = -4 \pi\alpha M_Z^2 \frac{F_V^2}{M_V^4} .
\label{TC3}\eq
Again, we have a linear constraint in the $(\delta_z, \delta_s,
\delta_{\gamma})$ space:
\bq
\delta_s = -\left(\frac{1-2\sss^2}{2\sss\ccc}\right) \delta_{\gamma} .
\label{TCrel}\eq
and the constraints
\bq
\delta_{z,s}>0 \ \ \ \ \ \ \delta_{\gamma}<0
\label{TCcons}\eq

Fig.\ref{189agc} shows, toghether with the projections on the three
coordinate planes of the allowed three-dimensional region 
for the $\delta$
parameters in the general case, the improved determination of
$\delta_z,~\delta_s$ and $\delta_{\gamma}$ achieved when the NP 
underlying the
$\delta$ parameters is assumed to be of AGC type. Techically, this
hypothesis cuts the three-dimensional region, in which the $\delta$ are
likely to lie, with a plane corresponding to the linear 
constraint in eq. 
(\ref{AGCrel}), thus effectively reducing the number of free 
parameters in the model to two. 
The data used here are $\sigma_\mu, \:\sigma_\tau, \: \sigma_5, \: 
{\rm A_{FB\mu}}$ and ${\rm A_{FB\tau}}$ measured by the four LEP
experiments at $\sqrt{s}=189$~GeV. 
The different contributions of the leptonic and hadronic cross sections 
to the combined constraints in the case of the two free parameter fit 
(AGC model) is shown in the plot by the enlarged contours 
which correspond 
to the result of the $\chi^2$ minimization 
when one of the two experimental 
contraints is released.  Since the measurements of $\sigma_5$ and
$\sigma_{\mu,\tau}$ provide in such model constraints almost
mutually orthogonal,
 and since the relative experimental uncertainty 
affecting the lepton 
forward-backward asymmetries is larger than that in the cross section
measurements, ${\rm A_{FB,\mu,\tau}}$ do not contribute 
significatively to
the overall contraint. 
The situation is different in the case on NP effects 
originated by the TC
model discussed above. The improved $1\sigma$ contours 
in the $\delta_z-\delta_s$
and $\delta_z-\delta_{\gamma}$ planes obtained when the TC model is considered
are shown in Fig.~(\ref{189tc}) (the same experimental inputs used for
the limits in Fig.~(\ref{189agc}) are used here). 
Although the statistical sensitivity of the test is not yet very
stringent\footnote{The contour displayed in Fig. \ref{189tc} for the TC
  model corresponds to a C.L. of 34\% for the values of two 
free parameters to fall
  simultaneously in the ellipse.}, one might incidentally notice that,
since this specific model of NP requires $\delta_z$ to be positive and
$\delta_{\gamma}$ negative, 
the determination of the $\delta$ parameters 
from the data does not seem to favour such a model.\par
At this preliminary stage, we should note that, while
no significative deviation from the SM prediction is observed, 
a systematic shift of the experimental results seems 
to appear in the three plans of pair of parameters
$\delta_{z,s,\gamma}$, with definite correlation signs.
Although this is not statistically significative,
it shows how our representation may reveal some features which are not
immediately visible on each observable separately. This is an
illustration of how it may be possible to get
some hints on NP properties.\par
The very successful operation of LEP in the recent 
years allows to foresee
that the final expected precision \cite{LibriGialli} 
on the observables measured in
two-fermion production processes will be reached and 
eventually overcome. 
Therefore, as a final work, we made the 
exercise of simulating the
final precision on the $\delta$ parameters that might be conceivably
achieved at the end of the LEP operation. 
With this aim, we assumed that the measurements of 
$\sigma_\mu, \:\sigma_\tau, \: \sigma_5, \: 
{\rm A_{FB\mu}}$ and ${\rm A_{FB\tau}}$ at $\sqrt{s}=183$~GeV and 
$\sqrt{s}=189$~GeV, with the experimental precision arising from the 
combination of the data from the four LEP experiments, can be 
complemented by four independent measurements of each variable 
with statistical precision corresponding to an integrated luminosity 
of $400~{\rm pb^{-1}}$ collected at an average center of mass energy 
of 200~GeV. 
No attempt of estimating the effects of systematic uncertainties, 
which might also be correlated between different measurements, 
experiments and center of mass energies, was made. 
Furthermore, since we just aim to estimate the final
contribution of LEP2 to the constraints on the parameterization 
of general NP model, we assumed that the final combined 
results will not show
deviations with respect to the SM. The width of the 95\% C.L. 
intervals on
each parameter (as obtained in a simultaneous fit) are listed in table
\ref{tab1} and the projections of the $\chi^2<\chi^2_{min}+1$
three-dimensional  region are shown in Fig.\ref{lep2final} 
and compared
with  the corresponding sensitivity derived from the data 
sets collected in
the past and already analised. 
The restricted domains allowed in the parameter space, when the specific
models of NP corresponding to AGC and TC are assumed, 
are also displayed 
for comparison.\par
One notes from inspection of Table \ref{tab1} that the available 
bound for the parameter 
$\delta_{\gamma}$ is more stringent than those for $\delta_z, 
\delta_s$.This is due to 
the fact that $\delta_{\gamma}$ contains the new 
physics effects on the photon 
exchange,
that is largely dominant in the muon cross section. 
The remaining parameters
$\delta_z, \delta_s$ contain the effect of the $Z$ 
exchange and of the $\gamma 
Z$
interference, and there is no unpolarized observable that 
priviledges one of
them. This would be the case of the longitudinal polarization asymmetry 
$A_{LR}$, 
that is drastically affected by the $\gamma Z$ interpherence 
and would therefore
represent a special test for $\delta_s$.\par
A final remark that can be added is that Table \ref{tab2},
which summarizes the 
correlations between 
the $\delta$ parameters in the general three parameter fit, 
suggests that other observables, affected by different combinations 
of the NP relic $\delta_i$, might be useful to disentangle to 
contributions of the different effects in a final analysis. 
A promising  candidate at LEP2 is the Bhabha scattering, to be
considered in a forthcoming paper.\par

Our analysis of possible effects of new physics is , at this point,
concluded. As explained in the Introduction, we have only considered
models that are {\it both} $\theta$ independent {\it and} of universal 
"smooth"
type, thus achieving remarkable simplifications, particularly in
our $Z$ peak subtracted approach where the existence of priviledged
$q^2$ values ($q^2=0$ and $q^2=M^2_Z$) 
can be usefully exploited (see Eqs.
(\ref{RUNP})-(\ref{DUNP})) to reduce the
number of parameters to a triplet of {\it constants}. There exist,
though, interesting models of new physics that do not meet all the
previous requests but nevertheless
involve a very small number of parameters, so that 
the analysis of their effects could
proceed in a simple way even in a "standard" treatment.
Using our approach, though, automatically takes into account one loop 
contributions (like e.g. $\Delta\rho$) already included in 
the used Z peak inputs. This is why, in this final part, we have retained
our approach and extended our analysis to the study of two simple models,
that we list here following the order in which they violate our
simplicity conditions.\par
a)Contact interactions.\\
The following interaction

\bq
\L={G\over\Lambda^2}\bar \Psi\gamma^{\mu}(a_e-b_e\gamma^5)\Psi
\bar\Psi\gamma_{\mu}(a_f-b_f\gamma^5)\Psi
\label{cont}\eq
\noindent
was first introduced with the idea of
compositeness \cite{contact}, but it applies to any 
virtual NP effect (for example higher vector boson exchanges)
satisfying chirality conservation 
(Vector and Axial Lorentz structures) and whose effective scale
$\Lambda$ is high enough so that one can restrict oneself to $dim=6$
operators.\par
The parameters $a_e,b_e,a_f,b_f$ can be
adjusted in order to describe all kinds of chiral couplings. For each
choice of pair of chiralities among 
L($a=b=1/2$), R($a=-b=1/2$), V($a=1,~b=0$),
A($a=0,~b=1$), there is only one free
parameter.

These models 
are not of universal type, but retain the property of being $\theta$
independent. Their contribution to all observables can be 
computed in our approach, by straightforward "projection"
on the \underline{four} $q^2$ dependent functions
$\tilde\Delta_{\alpha}$, $R$, $V^{\gamma Z}$, $V^{Z\gamma}$. 
By construction the model satisfies automatically
the parametrization of eqs.(\ref{RUNP})-(\ref{DUNP}), with three
constants 
$\delta_z$,$\delta^{\gamma Z}_s$,$\delta_{\gamma}$ 
and one new constant $\delta^{Z\gamma}_s$ that takes into account its
non universality, and have the expressions:

\bqa
&&\delta_{z,ef}=-({GM^2_Z\over\Lambda^2}){4\tilde s^2_l\tilde
c^2_lb_eb_f
\over e^2I_{3e}I_{3f}}\nonumber\\
&&\delta^{\gamma Z}_{s,ef}=-({GM^2_Z\over\Lambda^2})
{4\tilde s_l\tilde c_l(a_e-b_e\tilde v_l)b_f
\over e^2Q_eI_{3f}}\nonumber\\
&&\delta^{Z\gamma}_{s,ef}=-({GM^2_Z\over\Lambda^2})
{4\tilde s_l\tilde c_l(a_f-b_f\tilde v_f)b_e
\over e^2Q_fI_{3e}}\nonumber\\
&&\delta_{\gamma,ef}=({GM^2_Z\over\Lambda^2})
{(a_e-b_e\tilde v_l)(a_f-b_f\tilde v_f)
\over e^2Q_eQ_f}
\label{CT}
\eqa

Note that the appearance of the non universal extra quantity $\delta_{s, ef}^{Z\gamma}$
requires the use of the more complete 
Eqs.~(\ref{sig},\ref{U11pro},\ref{U12pro}) with the four functions
being expressed in terms of the $\delta_i$ through
Eqs.(\ref{RUNP})-(\ref{DUNP}).

b)Manifestations of extra dimensions.\\
Recently, an intense activity
has been developed on possible low energy effects of graviton exchange. 
The following matrix element for the 
4-fermion process $e^+e^-\to \bar{f} f$~\cite{graviton} is predicted:

\bq
{\lambda\over\Lambda^4}[\bar e\gamma^{\mu}e\bar
f\gamma_{\mu}f(p_2-p_1).(p_4-p_3)-\bar e\gamma^{\mu}e\bar f\gamma^{\nu}f
(p_2-p_1)_{\nu}(p_4-p_3)_{\mu}]
\eq

This model can be formally treated in our formalism in a way that is
analogous to that used in the previous case. 
One easily arrives, after a few steps, to the formal definitions:
 
\bqa
&&\delta_{z,ef}=-({\lambda M^2_Z q^2\over \Lambda^4})
{4\tilde s^2_l\tilde c^2_l
\over e^2I_{3e}I_{3f}}\nonumber\\
&&\delta^{\gamma Z}_{s,ef}=({\lambda M^2_Z q^2\over\Lambda^4})
{2\tilde s_l\tilde c_l\tilde v_l
\over e^2Q_eI_{3f}}\nonumber\\
&&\delta^{Z\gamma}_{s,ef}=({\lambda M^2_Z q^2\over\Lambda^4})
{2\tilde s_l\tilde c_l\tilde v_f
\over e^2Q_fI_{3e}}\nonumber\\
&&\delta_{\gamma,ef}=({\lambda M^2_Z q^2\over\Lambda^4})
{(\tilde v_l\tilde v_f-2cos\theta)
\over e^2Q_eQ_f}
\label{ED}\eqa

Comparing the above expressions with Eq.(\ref{CT}), one sees that
a first difference
is that now the four quantities $\delta_z$, $\delta^{\gamma Z}_s$,
$\delta_{\gamma}$, $\delta^{Z\gamma}_s$ are all proportional to $q^2$.
This is simply a kinematical manifestation of the higher dimension
of the Lagrangian, that shifts the lowest order coefficients of the
$q^2$ expansion to the first order in $q^2$ without
changing the philosophy of our approach.

The second difference is the appearance of a $\theta$ dependence in
the modification of the $\gamma$-coupling, as shown by Eq.(\ref{ED}).
This cannot be reabsorbed or eliminated, and forces one to 
restart from the more general 
Eqs.~(\ref{sig},\ref{U11pro},\ref{U12pro}) performing the full $\cos\vartheta$
integration.

In the final Table \ref{tab3} we have shown the present and (optimistic)
future LEP2 bounds on the two considered (a, b) models.
We have verified that a substantial agreement exists with the 
available ``standard'' calculations of the \underline{present}
bounds recently performed by LEP collaborations~\cite{LEPNPBounds}. This is
particularly encouraging since our approach is of a totally independent
nature.

A final comment is related to the weight of the different measurements
in the two considered cases (a,b). In the first model, the situation
that appears is essentially the same as in the universal examples that we
have considered, in the sense that the bulk of the information is 
provided by the two cross sections, with $\sigma_\mu$ still playing a 
major role. A totally different picture appears in the case of extra 
dimensions. Here, all the information is provided by the muon asymmetry.
The reason is not difficult to understand: the relevant parameter 
$\delta_\gamma$ is now proportional (neglecting the small $\tilde{v}_l^2$
term ) to $\cos\theta$. As a consequence of this fact, it only 
contributes $A_{FB, \mu}$ and not $\sigma_{\mu,5}$. This shows that, at 
least in the specific case of one interesting model, the role of the 
forward-backward muon asymmetry can become essential.

In conclusion, we have proposed a simple parametrization of the virtual
effects of a class of models of new physics in the present and future
four-fermion processes. This is particularly suited to treat the case
of effects of $\theta$-independent, smooth, universal type, but can be
easily extended to some cases of models that do not meet all previous
requests, by means of simple numerical programs that exist and are
available.

\newpage

\begin{table}[hbt]
\begin{center}
\caption{95\% C.L. bounds on $\delta_z,\:\delta_s$ and
$\delta_{\gamma}$ from 
the combined three free parameter fit of the general theoretical
expressions for $\sigma_\mu, \:\sigma_\tau, \: \sigma_5, \: 
{\rm A_{FB\mu}}$ and ${\rm A_{FB\tau}}$ to the present and future
measurements of fermion pair production at LEP2.  The 95\% C.L. bounds
obtained from a two free parameters fit to to the specific models of AGC
and TC, are also reported.  
\label{tab1}}
\vskip 0.5cm
\begin{tabular}{|c|c|ccc|}
\hline \hline
 Model  & DATA  & $\delta_z$  & $\delta_s$   & $\delta_{\gamma}$  \\
\hline \hline 
               &  {$\sqrt{s}=189$~GeV}      
& $-0.0027^{+0.036}_{-0.036}$ & $-0.0020^{+0.031}_{-0.031}$ 
& $-0.0026^{+0.0094}_{-0.0094}$ \\ 
3 free par.s   &  {$\sqrt{s}=183-189$~GeV}  
& $-0.0011^{+0.031}_{-0.031}$ &$-0.0033^{+0.027}_{-0.027}$ 
& $-0.0022^{+0.0081}_{-0.0081}$    \\ 
               &  Final LEP2       & $\pm$ 0.016 & $\pm$ 0.014 & $\pm$0.0043 
               \\ 
\hline \hline 
               &  {$\sqrt{s}=189$~GeV}      
& $-0.0014^{+0.0037}_{-0.0037}$ & $-0.0031^{+0.0074}_{-0.0074}$ 
& $-0.0026^{+0.0082}_{-0.0082}$ \\ 
         AGC   &  {$\sqrt{s}=183-189$~GeV}  
& $-0.0015^{+0.0032}_{-0.0032}$ &  $-0.0029^{+0.0064}_{-0.0064}$ 
& $-0.0022^{+0.0071}_{-0.0071}$ \\ 
               &  Final LEP2       & $\pm$0.0016 & $\pm$0.0033 & $\pm$0.0037
               \\ 
\hline \hline 
               &  {$\sqrt{s}=189$~GeV}      
& $-0.0061^{+0.015}_{-0.015}$ & $0.0014^{+0.0047}_{-0.0047}$ 
& $-0.0021^{-0.0075}_{-0.0075}$ \\ 
          TC   &  {$\sqrt{s}=183-189$~GeV}  
& $-0.0055^{+0.013}_{-0.013}$ & $0.0010^{+0.0041}_{-0.0041}$
& $-0.0016^{+0.0064}_{-0.0064}$\\ 
               &  Final LEP2       & $\pm$0.0066 & $\pm$0.0021 & $\pm$0.0034
               \\ 
\hline \hline 
\end{tabular}
\end{center}
\vskip 1cm
%\end{table}

%\begin{table}[hbt]
\begin{center}
\caption{Correlations between the best fit values of the $\delta$
 parameters obtained using the experimental constraints from 
$\sigma_\mu, \:\sigma_\tau, \: \sigma_5, \: 
{\rm A_{FB\mu}}$ and ${\rm A_{FB\mu}}$.  \label{tab2}} 
\vskip 0.5cm
\begin{tabular}{|c|c|ccc|}
\hline \hline 
     &       & $\delta_z$ & $\delta_s$ & $\delta_{\gamma}$ \\ 
\hline \hline
              &  $\delta_z$ & 1.00       & -0.94      & -0.01 \\ 
3 free par.s  &  $\delta_s$ & -0.94      & 1.00       &  0.28 \\ 
              &  $\delta_{\gamma}$ & -0.01      & 0.28       &  1.00 \\ 
\hline \hline 
         AGC  &  $\delta_z$ &  1.00      & 0.47      &    - \\ 
              &  $\delta_s$ &  0.47      & 1.00      &    - \\ 
\hline 
         AGC  &  $\delta_z$ &  1.00      &  -         &  0.12 \\ 
              &  $\delta_{\gamma}$ &  0.12      &  -         &  1.00 \\ 
\hline 
         AGC  &  $\delta_s$ &     -      & 1.00       &  0.93 \\ 
              &  $\delta_{\gamma}$ &     -      & 0.93       &  1.00 \\ 
\hline \hline 
          TC  &  $\delta_z$ & 1.00       & -0.89  & - \\ 
              &  $\delta_s$ &-0.89       &  1.00  & - \\ 
\hline 
          TC  &  $\delta_z$ & 1.00       &   -        &  0.89 \\ 
              &  $\delta_{\gamma}$ & 0.89       &   -        &  1.00 \\ 
\hline \hline 
\end{tabular}
\end{center}
\end{table}

\begin{table}
\begin{center}
\caption{
The following two tables show respectively the present and 
(optimistic) future LEP2 95\% C.L. bounds on the models (a,b)
as provided by using all measurements ($\sigma_l$, $\sigma_5$ and $A_{FB,l}$)
and by releasing one of the measurements. \label{tab3}} 
\vskip 0.5cm
\begin{tabular}{|c|cccc|}
\hline \hline 
${\Lambda}_{CT}$ (TeV)	& All 	& no $\sigma_l$ 	& no $\sigma_5$ 	& no $A_{FB,l}$ \\
\hline
&&&&\\
LL			& 2.9	& 1.8			& 2.8			& 2.9\\
RR			& 2.7	& 1.6			& 2.7			& 2.7\\
VV			& 4.7	& 2.7			& 4.6			& 4.6\\
AA			& 4.1	& 3.8			& 4.0			& 3.3\\
&&&&\\
\hline
${\Lambda}_{ED}$ (TeV)	& All 	& no $\sigma_l$ 	& no $\sigma_5$ 	& no $A_{FB,l}$ \\
\hline
&&&&\\
			& 0.78	& 0.78			& 0.78			& 0.25 \\
&&&&\\
\hline\hline
\end{tabular}

\vskip 0.5cm
\begin{tabular}{|c|cccc|}
\hline \hline 
${\Lambda}_{CT}$ (TeV)	& All 	& no $\sigma_l$ 	& no $\sigma_5$ 	& no $A_{FB,l}$ \\
\hline
&&&&\\
LL			& 4.0	& 2.5			& 3.9			& 3.9\\
RR			& 3.7	& 2.1			& 3.7			& 3.7\\
VV			& 6.4	& 3.6			& 6.3			& 6.3\\
AA			& 5.5	& 5.0			& 5.2			& 4.7\\
&&&&\\
\hline
${\Lambda}_{ED}$ (TeV)	& All 	& no $\sigma_l$ 	& no $\sigma_5$ 	& no $A_{FB,l}$ \\
\hline
&&&&\\
			& 0.89	& 0.89			& 0.89			& 0.30\\
&&&&\\
\hline\hline
\end{tabular}

\end{center}
\end{table}

\newpage 

\vskip 1cm
\section*{Figure Captions}

\begin{figure}
\vspace{18cm}
\includegraphics{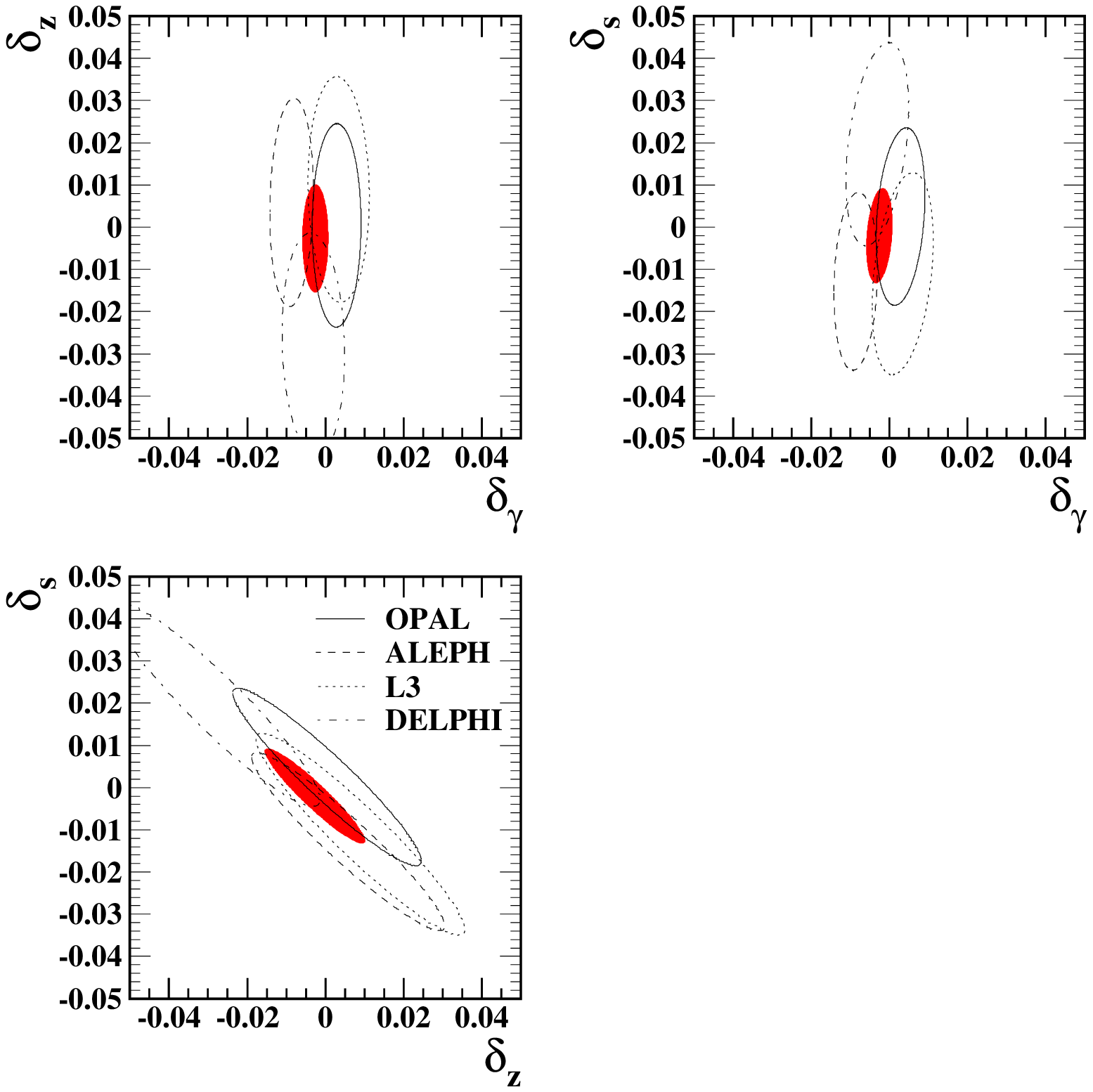}
\caption{\label{189_4exp}}
\end{figure}

{\bf Fig.\ref{189_4exp}:} 
Projections of the region in the $\delta_z,\delta_s,\delta_{\gamma}$ parameter
space defined by the condition $\chi^2<\chi^2_{min}+1$. The experimental
data used in the fit are the cross sections for quark, muon and tau pair
production and the forward-backward asymmetries of muon and tau pairs. 
The preliminary measurements, presented by the LEP collaborations at the 
EPS-HEP'99, obtained at $\sqrt{s}=189~$GeV have been used.  
The combination of the results of the four experiments (shaded area) is 
shown to sensibly improve the precision of the determination of the
$\delta$ parameters. 

\vskip 1cm

\begin{figure}[htb]
\vspace{18cm}
\includegraphics{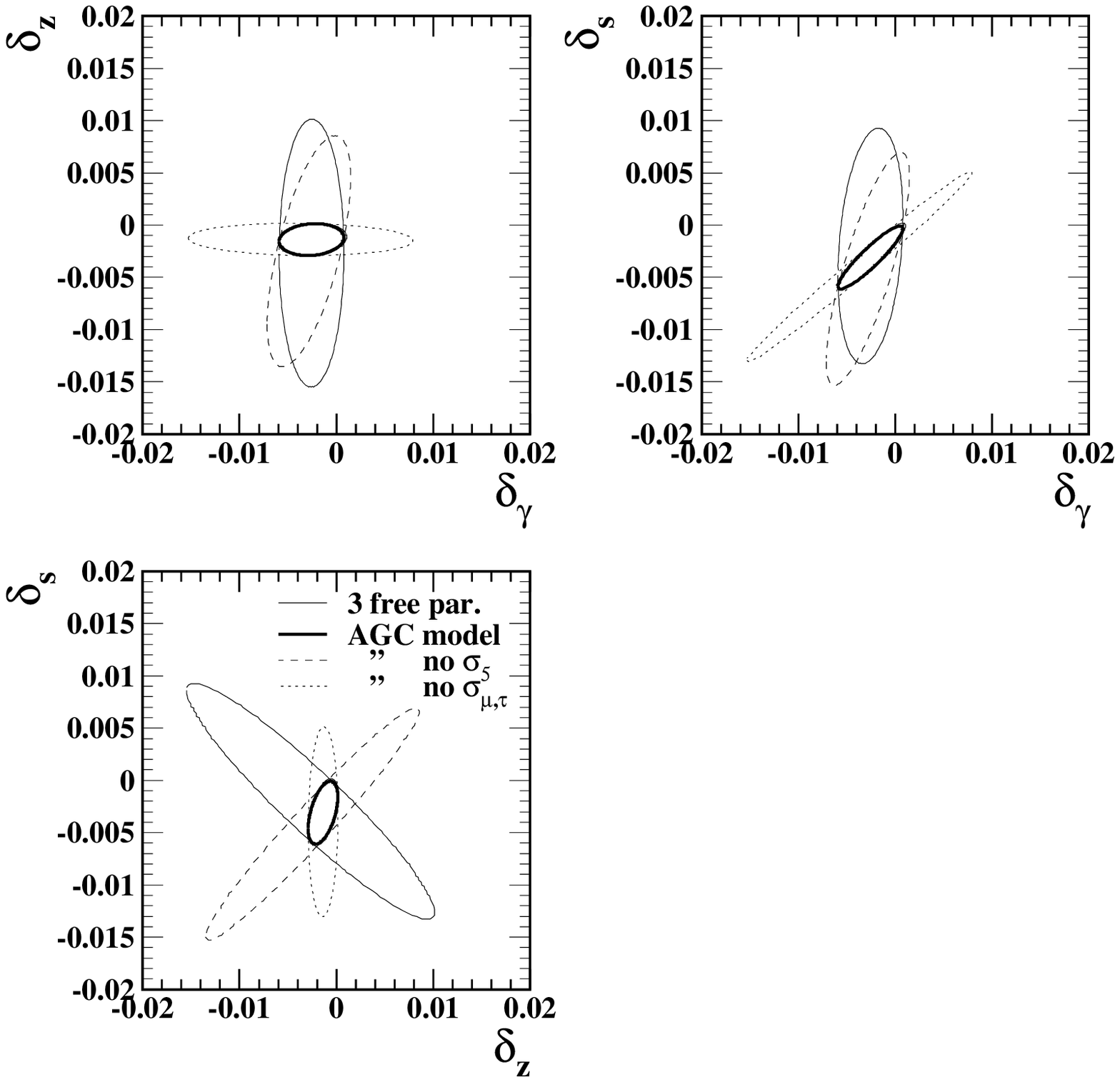}
\caption{\label{189agc}}
\end{figure}

{\bf Fig.\ref{189agc}:} 
Projections of the $\chi^2<\chi^2_{min}+1$ region in the
$\delta_z,\delta_s,\delta_{\gamma}$ 
parameter space obtained from the combined LEP results for 
$\sigma_\mu, \:\sigma_\tau, \: \sigma_5, \: 
{\rm A_{FB\mu}}$ and ${\rm A_{FB\tau}}$ at $\sqrt{s} = 189$~GeV. 
The small ellipses represent the projections on each plane of the 
intersection between the three dimensional ``allowed region for the 
$\delta$ parameters'' and the plain corresponding to the linear 
constrain between them, arising from the AGC model discussed 
in the text.

\vskip 1cm

\begin{figure}[htb]
\vspace{11cm}
\includegraphics{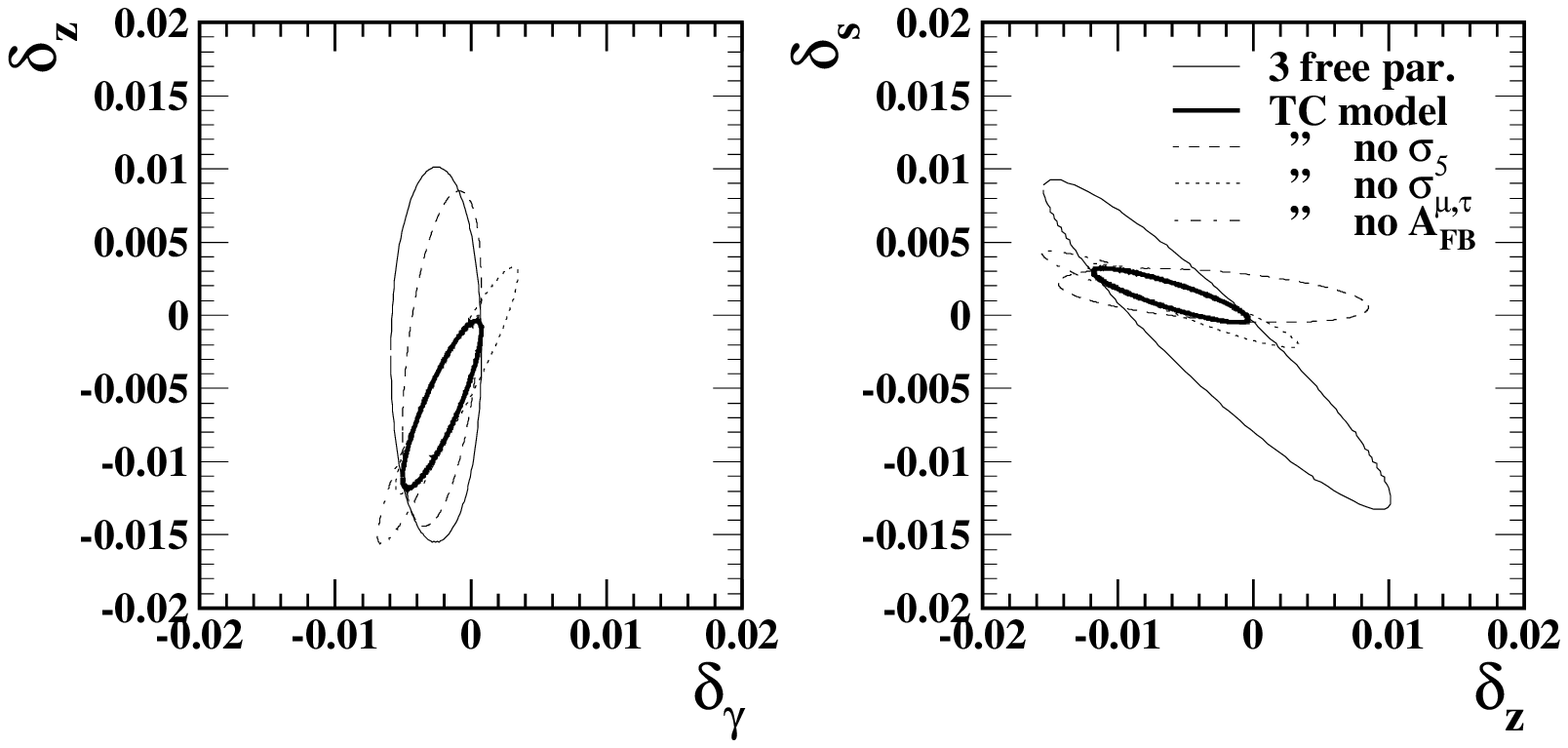}
\caption {\label{189tc}}
\end{figure}

{\bf Fig.\ref{189tc}:} 
The reduction of the allowed parameter space for
the $\delta$ parameters in the hypothesis of virtual effects driven by 
Technicolor type of New Physics is shown here.  
The definition of the projected ellipses and the experimental constraints 
applied in the $\chi^2$ are as in Fig.~(\ref{189agc}).  

\vskip 1cm

\begin{figure}[htb]
\vspace{18cm}
\includegraphics{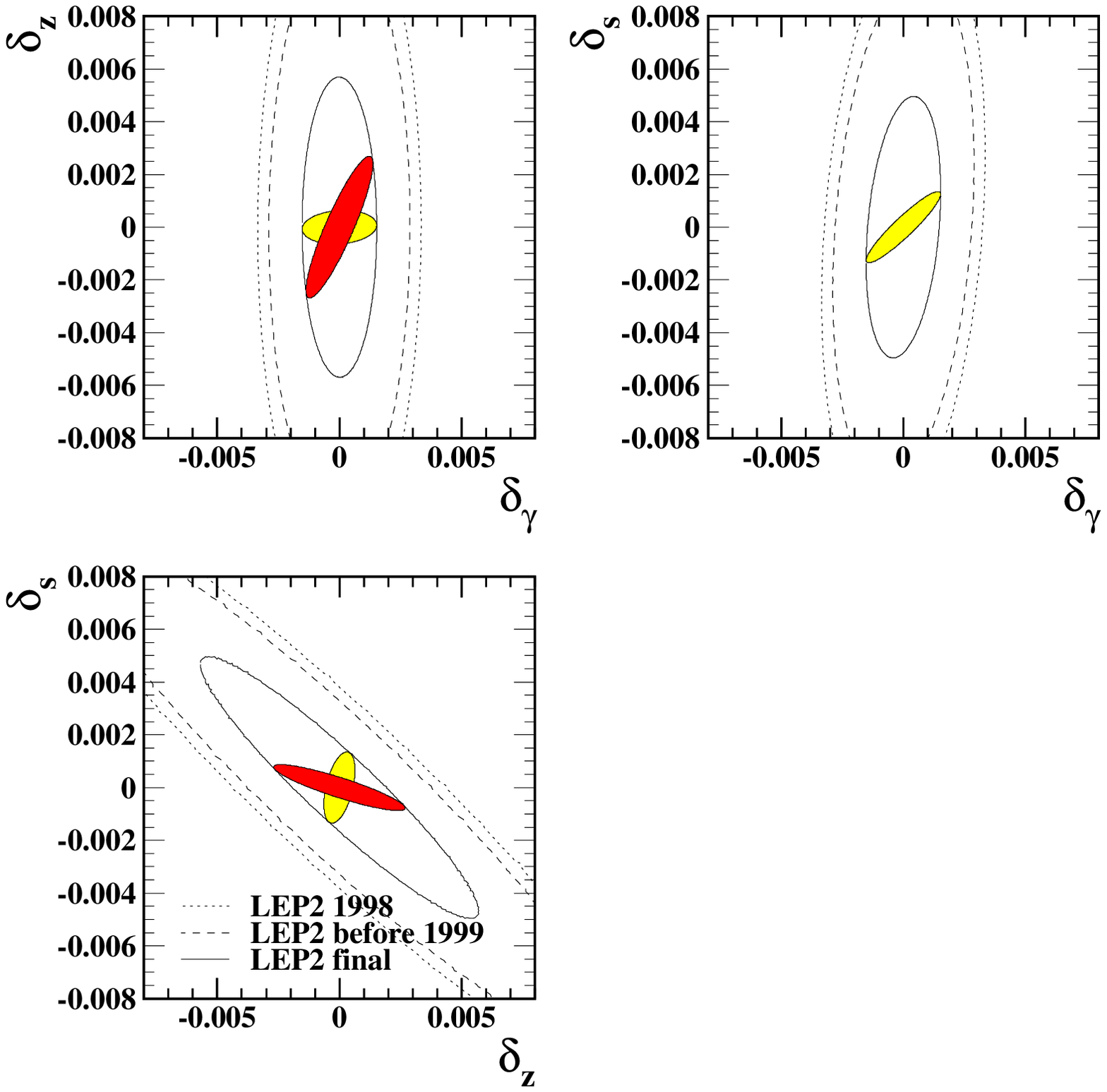}
\caption{\label{lep2final}}
\end{figure}

{\bf Fig.\ref{lep2final}:}
Envisaged final constraints on the $\delta$ parameters from the whole 
LEP data set. The contours, corresponding to $\chi^2=\chi^2_{min}+1$, are
derived by assuming that the present available measurements of 
$\sigma_\mu, \:\sigma_\tau, \: \sigma_5, \: 
{\rm A_{FB\mu}}$ and ${\rm A_{FB\tau}}$ at 183~GeV and
189~GeV might be complemented by analogous measurements from a total
integrated luminosity of 400~$\rm pb^{-1}$ collected at an average center of
mass energy of 200~GeV. Statistical uncertainties only have been assessed 
on the future measurements and agreement of the final combined results 
with respect to the Standard Model predictions has been assumed. 
For comparison the sensitivity achieved with the currently abailable data
is shown. The light and dark shaded areas represent the final constraints 
in the AGC and TC models.

\newpage

\newpage

\newpage

\newpage

\end{document}